# A Data Ecosystem to Support Machine Learning in Materials Science

Ben Blaiszik[1,2,†], Logan Ward[1,2], Marcus Schwarting[2], Jonathon Gaff[1], Ryan Chard[1,2], Daniel Pike[3], Kyle Chard[1,2], Ian Foster[1,2,†]

[1] University of Chicago

[2] Argonne National Laboratory

[3] Cornell University

[†] Corresponding authors: Ben Blaiszik (blaiszik@uchicago.edu), Ian Foster (foster@anl.gov)

**Abstract (100 words):**

Facilitating the application of machine learning to materials science problems requires enhancing the data ecosystem to enable discovery and collection of data from many sources, automated dissemination of new data across the ecosystem, and the connecting of data with materials-specific machine learning models. Here, we present two projects, the Materials Data Facility (MDF) and the Data and Learning Hub for Science (DLHub), that address these needs. We use examples to show how MDF and DLHub capabilities can be leveraged to link data with machine learning models and how users can access those capabilities through web and programmatic interfaces.

# 1  Introduction

A growing opportunity exists for the materials science community to leverage and build upon the advances in machine learning (ML) and artificial intelligence (AI) that are reorienting and reorganizing industries across the economy. In materials science, there is well-founded optimism that such advances may allow for a greatly increased rate of discovery, development, and deployment of novel materials, bringing researchers closer to realizing the vision of the Materials Genome Initiative [1]. However, despite considerable growth in the number of materials datasets and the volume of data available, researchers continue to lack easy access to high-quality machine-readable data of sufficient volume and breadth to solve many interesting problems. They also struggle with growing diversity and complexity in the data science and learning software required to apply ML and AI techniques to materials problems: software that includes not only materials-specific tools but also a wide range of other data transformation, data analysis, and ML/AI components, many not designed specifically for materials problems. Seizing the opportunity of ML and AI for materials discovery thus requires not just more and better data and software: it requires new approaches to navigating and combining data sources and tools that allow researchers to easily discover, access, integrate, apply, and share diverse data and software.

We describe in this article two related materials data infrastructure systems that address these needs: the Materials Data Facility (MDF) [2] and the Data and Learning Hub for Science (DLHub) [3]. MDF serves as an interconnection point for materials data producers and consumers. Its services allow data to be collected from many sources, to be enriched with a variety of tools (e.g., automated metadata extraction, quality control), and to flow onwards to many destinations, including not only MDF-operated services (e.g., the MDF Publish repository, for storage of data with no other home, and the MDF Discover search index, for integration, navigation, and search of any and all data known to MDF), but also to the growing number of other materials-related data infrastructure components (e.g., 4CeeD [4], Citrination [5], NIST Materials Resource Registry [6]). DLHub provides similar functions for ML/AI models and associated data transformation and analysis tools, allowing researchers to describe and publish such tools in ways that support discovery and reuse; run published tools over the network (with tools executed on a scalable hosted infrastructure); and link models, other tools, and data sources into complete pipelines that can themselves be published, discovered, and run.

In the sections that follow, we briefly review the state of the materials data ecosystem; describe MDF and DLHub goals, features, and service architectures; and present three examples that showcase how MDF and DLHub can be applied to materials science problems. We conclude with thoughts on future directions for the materials data ecosystem.

## 2  The Materials Data Ecosystem

Many types of tools are available to address myriad problems in handling materials data. Most prominent are the extensively-curated and specialized data repositories of materials data, including high-throughput density functional theory (DFT) databases [7–11] and polymer property databases [12,13]. Tools like Citrination [5] and the Configurable Data Curation System (CDCS-formerly MDCS) [6] allow scientists to quickly create and share new databases. Curated databases provide data in well-structured forms that are immediately-accessible to data analysis software, but only represent a small fraction of the useful materials data.

General-purpose publication repositories (e.g., NIST Materials Data Repository [6], Zenodo, Figshare) provide the ability for researchers to make data available to others, even if those data have not yet undergone the extensive curation typically needed to produce structured datasets. Laboratory Information Management System (LIMS) and workflow management tools like 4CeeD [4] and Materials Commons [14] provide a route for curating data and tracking provenance as data are produced. Together, these tools offer a rich environment of data ready for use in materials research.

As data availability has increased, a concomitant growth has occurred in software tools to simplify and automate common tasks in the materials informatics pipeline: for example, the MAterials Simulation Toolkit for Machine Learning (MAST-ML) [15], the Materials Knowledge System in Python (pyMKS) [16], matminer [17], pymatgen [18], and the Atomic Simulation Environment (ASE) [19]. Another critical community effort is the NIST Materials Resource Registry (MRR) [6], a federated set of registries built to enable registration and discovery of datasets, software, projects, and organizations relevant to materials science. Together, these tools, data services, and software comprise many of the components needed to speed the application of materials informatics and ML.

## 3 Materials Data Facility (MDF)

While the materials data ecosystem described previously has grown considerably in the volume of data available and the number of available tools, there remain many opportunities to enhance the value of individual components by connecting them in ways that leverage and maximize their unique strengths. Such connections would enable a thriving materials ecosystem in which new data gathered at any repository are automatically dispatched to other repositories; new services are easily constructed from a growing set of modular software and service components; new service capabilities are applied automatically to appropriate data streams; and new machine learning studies are easily bootstrapped from data gathered with a single query from dozens of repositories, and analyzed with models from multiple sources.

MDF supports this vision by providing an interconnection point that allows producers of materials data to dispatch their results broadly and data consumers to discover and aggregate data from independent sources. It streamlines and automates data sharing, discovery, access, and analysis by: 1) enabling data publication, regardless of data size, type, and location; 2) automating metadata extraction from submitted data into MDF metadata records (i.e., JSON formatted documents following the MDF schema [21]) using open-source materials-aware extraction pipelines and ingest pipelines; and 3) unifying search across many materials data sources, including both MDF and other repositories with potentially different vocabularies and schemas. Currently, MDF stores 30 TB of data from simulation and experiment, and also indexes hundreds of datasets contained in external repositories, with millions of individual MDF metadata records created from these datasets to aid fine-grained discovery.

### 3.1 Collecting and Sharing Data with MDF

The diversity and scale of materials data can pose challenges for both data producers and data consumers. Data producers can find it difficult to determine which repository best suits their dataset, or they may have specialized requirements (e.g., support for large datasets, advanced curation flows, or varying access control across the dataset lifecycle) that are not met by any single repository. Consumers attempting to locate data face yet more difficulties, as in order to collect data, they must often first navigate differing web and programmatic interfaces, and then merge data cataloged and described by different metadata schemas. MDF takes important steps towards addressing these challenges by supporting collection of data from many locations, enriching and transforming those data in materials-aware ways, and managing interactions with many data services.

MDF consists of three modular services: MDF Publish, MDF Discover, and MDF Connect. MDF Publish is a decentralized dataset repository. It allows a user to publish a dataset to any Globus endpoint [20], identify the published dataset with a permanent identifier, and implement user-driven dataset curation flows. MDF Discover provides a scalable, access-controlled, cloud-hosted, materials-specific search index, coupled with software tools to enable advanced user queries and data retrieval. MDF Connect is the central element that connects not only MDF Publish and Discover, but also external services (Figure 1). It supports three primary actions: 1) **submission** via user requests from programmatic or web interfaces triggers the MDF Connect service to collect from many common sources; 2) **enrichment** of collected data through general and materials-specific metadata extraction, combination of extracted and user-provided metadata into MDF metadata records, and transformation of dataset contents (e.g., from proprietary to open formats); and 3) **dispatch** of data to MDF Publish, metadata to MDF Discover, and combinations of data and metadata to other community data services selected by the user. MDF Connect can collect data from cloud storage providers (Google Drive, Box, Dropbox), distributed storage systems accessible via Globus [20], community repositories (NIST Materials Data Repository [6], Figshare, Zenodo), and materials-specific data services (e.g., 4CeeD [4]).

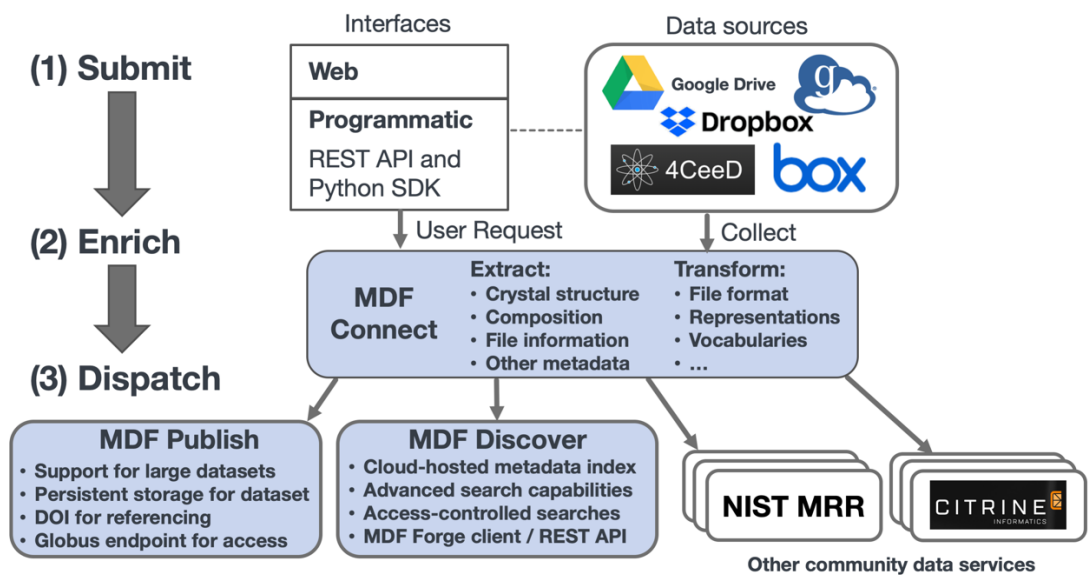

Figure 1: Materials Data Facility (MDF) overview. (1) Users submit data to MDF by specifying the data's location, title, authors, and more. (2) MDF Connect collects data from the specified location and applies materials-specific extractors and transformations to enrich the data. (3) Processed data and metadata are dispatched to any supported community data service(s) specified by the user. Other users can then discover, interact with, and access the data using any of those services.

As described earlier, when a user submits a dataset to MDF Connect, the MDF Connect service collects the data, enriches the data by extracting further information from the files, and dispatches that information to the wider ecosystem. MDF Connect enriches data by invocation of a series of extractors that extract general information (e.g., file name and size) and scientific information (e.g., crystal structure or material composition) from the files provided by the user to facilitate discovery. Subsequently, this extracted information is merged with user-provided metadata to create an MDF metadata record following the MDF schema [21]. For example, MDF Connect is able to extract the inputs and outputs of electron microscopy images and DFT codes, among others, to generate descriptions of the material being studied, identify instrument settings, or capture computed properties and convert these into MDF metadata records that are dispatched to the MDF Discover index. Advanced users can build new data extractors for the ecosystem or provide additional information to help MDF Connect make sense of the data provided (e.g., by providing a mapping between fields in a custom comma-separated values file and the MDF schema [21]). MDF Connect also allows optional dispatch of this information to other host services, such as Citrination and MRR. Thus, with minimal effort from the researcher, MDF Connect provides enriched descriptions of the submitted data and makes it available in many forms.

### 3.2 Data Discovery with MDF

The rapidly growing quantities of data contained within both MDF Publish and other community repositories makes discovery of datasets based on their attributes or contents a challenging problem. No single schema can cover all data types, yet users want to be able to search across these diverse data. MDF Discover addresses these needs by operating a flexible search index in which registered datasets and associated files are described by key-value pair metadata records (e.g., JSON documents) created by MDF Connect that follow a metadata schema extending the

DataCite and NIST Materials Resource Registry conventions [21]. For added flexibility, MDF Discover also allows for addition of up to 10 user-defined metadata fields per dataset on which searches can be performed. This search index is operated in the cloud for scalability and availability, with a REST API permitting both programmatic and web access.[22]

MDF Discover aims to allow simple data discovery while also permitting advanced querying and data faceting when required. To this end, it implements a query syntax that supports full-text matching (i.e., matching of query text against the value of any key in the registered metadata), direct querying against user-specified keys, typed range queries for dates and numeric fields, fuzzy matching, and wildcard matching. Additionally, users may discover data through faceting operations that allow users to retrieve summary statistics, partition matching data into buckets, and drill down into these buckets with subsequent queries. For example, a user may facet data by elemental composition to determine how many records are available for different elements, and then select a single element of interest to investigate further.

Another important feature of MDF Discover is the ability to define access controls on each registered record. While most data and metadata registered with MDF are publicly accessible, access control mechanisms allow users to define which users or groups of users can access certain metadata. Multiple metadata records may be associated with a single dataset or the contained files, with differing permissions allowing for different users to see different views depending upon the user permissions. These capabilities promote an open, participatory environment where users can contribute to the description of a dataset incrementally, with assurance that only authorized users can see metadata records until they are ready for broader sharing. When a user searches MDF, their search results reflect only the metadata records that they are authorized to access.

Many materials scientists interact with data, and share analysis methods, by writing Python programs. To support these users, MDF provides the MDF Forge Python client to make it easy to write Python programs that use MDF Discover capabilities to perform common search and data collection tasks, such as searching by dataset name, author names, or elemental composition, and that then use Globus or HTTPS methods to retrieve data records identified by such searches. For example, Figure 2 shows a user first querying by dataset name (i.e., source_name) and elemental composition and then retrieving results, including all referenced files, with Globus. A similar data retrieval mechanism allows MDF Forge users to retrieve data by HTTPS, although this route can be significantly less performant for data aggregations that include large files or many files. Users can install MDF Forge on various operating systems via the Python Package Index (PyPI).

```python
from mdf_forge.forge import Forge
mdf = Forge()

res = mdf.search_by_elements(['Co','V'], source_names=["oqmd"])
mdf.globus_transfer(res) # Defaults to local Globus endpoint
```

Figure 2: Example showing a query and data retrieval operation using the MDF Forge Python client. A Forge Python client is instantiated, and a search is performed to find records from the Open Quantum Materials Database (OQMD) that contain cobalt or vanadium. The result set is then passed to the globus_transfer function to transfer the files associated with the matching results, including the full simulation output for each record, to the user's local Globus Connect Personal endpoint.

## 4  The Data and Learning Hub for Science (DLHub)

A key factor slowing the adoption of data-driven materials science approaches is that few machine learning models and other related codes developed by materials scientists are easily accessible. Even when open-source codes are shared via mechanisms such as GitHub, they can be difficult to install, train, and run. The commercial use of machine learning has benefited from the availability of web interfaces that provide simple routes for employing common machine learning tasks (e.g., the Rekognition tools from Amazon). Similar capabilities are required for research models.

We created DLHub to make it possible for scientists to make models accessible via web interfaces with minimal effort. Much as MDF connects data providers and consumers across the materials science community, DLHub [3] connects data with reusable data transformation and model serving capabilities, allowing producers of such capabilities to make them easily available, and permitting consumers to quickly discover the latest AI/ML developments and to apply those developments to their research projects. Thus, for example, a researcher working with scanning transmission electron microscopy images may be able to easily discover and apply a model to assess image quality or to detect loop defects.

DLHub seeks to overcome inefficiencies in the ML life cycle by providing facilities that allow researchers to describe, publish, discover, and run ML models and associated data transformation and analysis codes with minimal overhead. Using DLHub, a researcher can discover, for example, that a model exists for prediction of materials structure and phase from x-ray coherent diffraction data (see Section 5.2), that this model is accessible at a persistent DLHub web address (URL), and that the model expects as input images of shape 32x32 pixels and produces as output images of shape 32x32 pixels. The researcher can then use the DLHub SDK to call this model from any network-connected computer without needing to download, configure, and run the model on their local PC. By eliminating the need to install complex software, DLHub greatly reduces the overheads associated with reusing and programmatically incorporating models and other software into analyses, services, or other code.

Under the hood, the DLHub service organizes user-supplied metadata and software (models, custom functions, etc.) to create portable and scalable Docker container *servables*; registers servables along with descriptive metadata in a catalog to support discovery; and deploys servables onto scalable computing systems to permit rapid execution in response to user requests. Chard et al. [3] provide a full description of the DLHub service and architecture.

## 4.1 DLHub Capabilities

The DLHub service and SDK allow researchers to perform four key actions (see Figure 3): 1) **describe** a software tool by providing the information that the DLHub service needs to create and to permit discovery and use of the associated servable; 2) **publish** a servable, sending files and metadata to DLHub to create the servable and register it in the servable catalog; 3) search the catalog to **discover** interesting servables and to learn how to use them; and 4) **run** a servable against provided input data on DLHub-provided computing infrastructure. Users can access each of these capabilities through the DLHub SDK, a REST API, or a command line interface (CLI).

**Describe**: A researcher first uses DLHub SDK functions to specify the models, code, and data (e.g., neural network weights) that will be used to construct a new servable, and to provide descriptive metadata such as author names, links to source code, input types and shapes, and domain of application. The initial DLHub schema leverages and extends work from Kipoi [23] and DataCite [24]. This information allows other scientists to determine whether the servable is applicable to their problem and what information to provider when invoking it. The DLHub SDK provides utilities that, for common types of servables (e.g., Keras, Tensorflow, ScikitLearn, and Torch, model objects), can extract automatically much of the information needed to describe and recreate the servable (e.g., input and output shapes, neural network architecture, locations of important files, software dependencies). Because scientific software often involve custom libraries, DLHub is not limited to only common libraries and can serve arbitrary Python functions.

A researcher who describes a model to DLHub can also specify whether access to the created servable and its metadata is to be fully open or, alternatively, constrained to a group or a defined list of users. Access policies for a servable and its metadata can be different: thus, for example, metadata may be open, permitting discovery, while access to the servable itself constrained to specified individuals. A user can thus discover that a servable exists and proceed to request access.

**Publish**: A user can then send a request to DLHub to publish the servable. A publish request collects the metadata and files specified when the user describes their model and dispatches this information to the DLHub service. Upon receiving the model description, the DLHub service builds the servable into a portable container (e.g., Docker or Singularity) and loads the combined metadata into the servable catalog. Note that the container is constructed automatically by the DLHub service; the user need not install any software on their system for that purpose. Once the build process is complete, a unique DLHub service endpoint is constructed to which users can send requests to invoke the newly published model. Optionally, a user can choose to associate a unique identifier, like a digital object identifier (DOI), with the servable to enhance citability.

**Discover**: Users can discover published servables, for which they are authorized to view metadata, by querying the associated metadata. The full set of fields are described in the DLHub schema which can be found from the DLHub web page linked in the Code and Data Availability section. A servable query, like the data queries used by the MDF (Section 3), can combine full-text search, range queries, pattern matching, and wildcards. Query results provide added context about how to use the servable(s) located. For example, the user might learn that the model accepts images of a specific size, find a link to associated journal articles, or be directed to a test set of data they can use with the servable. This functionality is accessed through the DLHub SDK, CLI, or a web interface that is under development.

**Run:** Once a user has found a servable that they want and are authorized to run, they can invoke it on input data that they supply, and receive the resulting output in response. The servable itself runs remotely on services that execute on a 200-processor cluster at Argonne National Laboratory, Amazon Web Services, or other supported resources. Requests can be synchronous (i.e., the reply from DLHub service contains the outputs) or asynchronous (i.e., the reply contains a key used to retrieve the results later). These two modes allow DLHub to support servables with both fast and slow execution times.

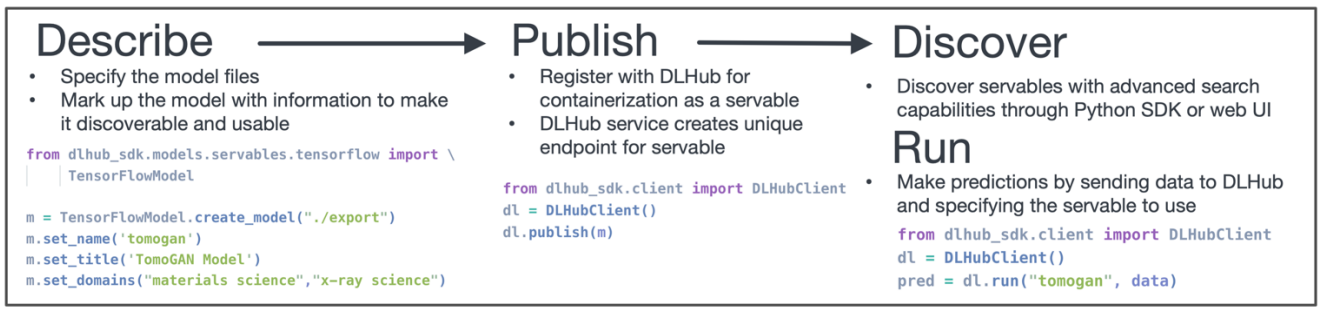

Figure 3: DLHub usage pattern. A user describes a model or custom code to make it discoverable and reusable. The user submits a publish request to DLHub, triggering a transfer of the required files and registration of the model metadata with the DLHub service. A servable is created to hold the model and custom code, and the associated metadata are loaded into a registry for later discovery. Subsequently, users can run the servable with new inputs to receive the defined output from the servable.

## 4.2 DLHub Service

The DLHub service is designed to support the publish, discover, and run capabilities just discussed with high availability, scalability, security, and performance. (The describe capability is handled by the DLHub SDK.) To this end, the DLHub service's implementation comprises multiple components: a cloud-hosted DLHub service that accepts user publish, discover, and run requests; a cloud-hosted servable creation service; a cloud-hosted metadata catalog used to serve discover requests; and potentially many servable execution environments that are able to run on cloud resources, Kubernetes clusters, high-performance computing systems or elsewhere.

DLHub packages servables as Docker or Singularity *containers*, a computing technology that facilitates portability across different computing resources and provides a sandboxed execution environment. DLHub uses Parsl [25] to manage scalable computing resources by deploying servable containers on provisioned nodes or cloud instances. It sets up a high-performance connection between the DLHub service and the deployed servable. When a user executes a servable, the DLHub service retrieves the servable metadata and validates the input parameters. The request is packaged with servable metadata and serialized for transmission to an execution node capable of running the servable; the servable is deployed on one or more nodes; the DLHub service transmits the input data to the servable for execution; and the servable sends results back to the DLHub service when complete.

# 5 Science Use Cases

To illustrate the power of these approaches, and how DLHub and MDF capabilities can be included in scientific workflows, we describe three science use cases. The first combines data hosted and indexed within MDF Publish and Discover, respectively, with ML models published as servables to DLHub to rapidly predict band gap based on an input image. The second uses a DLHub servable to extract the corresponding material structure from a set of input X-Ray coherent diffraction images. The third uses several different DLHub servables to make more accurate atomization energy predictions using low fidelity simulations as input, with the training and test data indexed and hosted in MDF Publish and Discover. Code for each example is accessible as described in the Code and Data Availability section.

## 5.1 Combining DLHub and MDF to Facilitate Band Gap Prediction

Stein et al. [26] recently described ML methods for predicting material band gap and spectra from optical images obtained experimentally. The dataset used to train their ML model contained optical absorption spectra and optical images of samples prepared via high-throughput techniques [27]. They also described methods for training a variational autoencoder (VAE) model and for training an ML model on the resultant encoder latent space to predict a sample's optical absorption spectrum. The band gap can then be determined from these predicted spectra by using the multiple adaptive regression splines (MARS) algorithm to locate the absorption onset [28].

To make this work easily reusable, we submitted the training dataset to MDF Connect, dispatched the extracted metadata in MDF Discover, and published the models referencing those data in DLHub. The Stein dataset consists of MDF metadata records describing 180,902 individual optical images and output spectrum files, so as to enable creation of different training and test datasets by sub-selecting data by index or materials composition as needed. The experimental dataset used in this study had previously been available only as a single, large HDF5 file. Now, with MDF, the dataset can be partitioned via user queries, immediately enabling new applications and data mixing opportunities. Second, we published several models into DLHub (as servables) based on, and extending, the models of Stein et al. [26], including a first model that resembles the optical image VAE described in the paper, a second optical image autoencoder (AE) model, and a third model that instead uses color clustering techniques to predict the material bandgap. We then used the dataset, as available in MDF Discover, to streamline the process of retrieving data (Figure 4a) and running servables within DLHub on the retrieved data (Figure 4b). We demonstrated autoencoding of the original images (Figure 4c) and examination of the latent space of the trained VAE (Figure 4d) with respect to image color and bandgap, all with only a few lines of code.

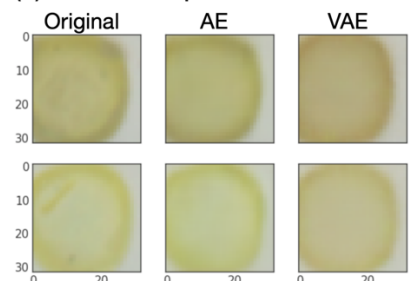

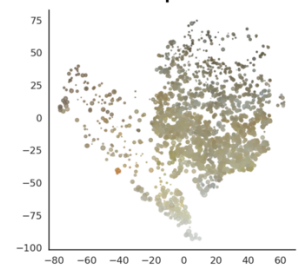

Figure 4: Retrieving test data through MDF and passing data to DLHub for prediction of absorption onset (band gap). (a) Using the Forge Python client, the input image and absorption spectrum are aggregated from the stored dataset. (b) The retrieved data are sent as input to the encoder model in DLHub to produce a latent representation of the image. (c) Two original input images and corresponding AE and VAE model outputs. (d) A sampling of the VAE latent space clustered by the TSNE algorithm [29], with each circle color representing the image color and with size proportional to band gap for a subset of the data. Links to the code to reproduce this work can be found in the Code and Data Availability section.

## 5.2 Coherent Diffraction Imaging Prediction

X-Ray coherent diffraction imaging (CDI) is an experimental technique that allows for determination of material structure and phase, with the phase encoding many interesting material properties, such as strain state [30]. To enable rapid phase and structure predictions from CDI data, Cherukara et al. [30] built a deep convolutional neural network to predict phase and structure, using a set of simulated CDI images of varying structures with varying strains for model training. Given the importance of CDI in materials science, especially at the Advanced Photon Source at Argonne National Laboratory, the widespread availability and deployment of such a model would be of great value, enhancing the ability to gather information quickly on samples and to assess the state of an experiment in a control loop.

In this case, we first submitted the available test data to MDF. Next, as the trained model is freely available via GitHub, we simply described it with metadata to credit the authors and enhance discovery, and then published it into DLHub using the DLHub Python SDK to pass the model's GitHub location and metadata to DLHub. The DLHub service then collected the model files, here comprising a set of Keras saved weights and the model architecture, and created a servable and an associated servable endpoint automatically. Researchers can thus test and run this servable on data with minimal coding knowledge or software installation overhead. In fact, only three lines of code, as shown in Figure 5(a), in addition to the data retrieval code (omitted for brevity) are needed to run this servable. Two examples of an input CDI and the predicted output structure are shown in Figure 5(b).

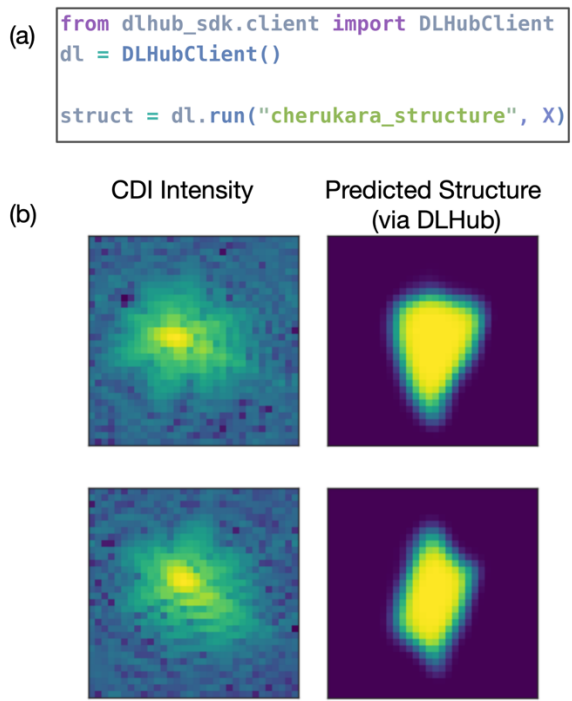

Figure 5: Predicting material structure based on CDI data, via DLHub. (a) Using the DLHub Python client, the input X-ray diffraction data (32x32 pixel CDI intensities) are run with the cherukara_structure servable in DLHub. (b) Example CDI intensity (left) input and (right) predicted structures from the DLHub servable. Links to the code to reproduce this work can be found in the Code and Data Availability section.

## 5.3 Fast, High-Quality Estimates of Molecular Atomization Energies

The ability to predict the energy of a molecule accurately from first-principles calculations forms the core of many approaches for the discovery and rational design of materials. However, while *ab initio* methods exist that can predict the energy of molecules with accuracies comparable to the uncertainty in corresponding experimental data (e.g., G4MP2 [31]), the computational expense of these high-accuracy methods limits their widescale use. To address this issue, Ward et al. [32] built machine learning models that use a recently-published MDF dataset to predict high-accuracy, G4MP2 energies from the outputs of faster, but inaccurate calculations (B3LYP). The authors used DLHub to make this capability available to the wider community.

Ward et al. [32] used well-established techniques from the materials literature to build this capability. The models were trained using a deep learning approach by Schütt et al. (SchNet [33]) to learn the differences between B3LYP- and G4MP2-level atomization energy calculations (i.e., Δ-learning [34]). The best resulting models predict G4MP2-level molecular atomization energies with an accuracy of around 10 meV, far below the difference between G4MP2 and experimental values. The models produced in this work are available in DLHub so that others in the community can use them without needing to install any software. As shown in Figure 6, it is possible to correct the B3LYP atomization energies for hundreds of molecules per second by using the model from Ward et al. through DLHub – putting the ability to quickly estimate high-fidelity atomization energies in the hands of any molecular modeler. This capacity can be scaled elastically with demand by increasing or decreasing the number of servable replicas running in DLHub.

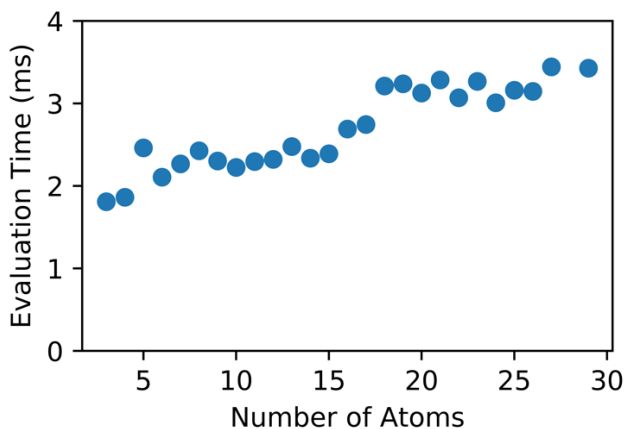

Figure 6: Time to predict the G4MP2-level atomization energy of 100 molecules given input B3LYP energies and relaxed structure using a machine learning model produced by Ward et al. [32] as a function of molecule size. Timings were measured over 64 identical runs with one servable container running in the DLHub service. Throughput may be scaled elastically by varying the number of container replicas. Error bars representing the standard error of the mean are within the size of the markers. Links to the code to reproduce this work can be found in the Code and Data Availability section.

# 6  Conclusion

We have described key MDF and DLHub capabilities that we argue are critical to building a materials data ecosystem that is optimized to enable the widespread application of machine learning and artificial intelligence methods. These capabilities include automation of data sharing (even for large datasets) among heterogenous data services; enrichment of data with both general and materials-specific metadata to promote discovery and reuse; software tools to simplify data discovery, aggregation, and use; and a library of curated machine learning models and processing logic that can easily be applied to new data streams. Each capability is provided as a service, greatly reducing the work required for users to access these powerful capabilities and enabling other data services to leverage them in a modular fashion. We presented three examples to showcase some of the many ways in which MDF and DLHub capabilities can be used to deliver the results of machine learning studies in materials science to a broad audience.

In future work, we plan to build connections between MDF and other data services to encourage broader and more simple dissemination and discovery of datasets. Towards improving data discovery, we see a clear opportunity for various projects to combine efforts to collaboratively build the software necessary to automate the extraction of metadata from hundreds of common file types used in materials science, since much of this work is currently fragmented across several code bases. Further, we are encouraged to see that many data repositories now enable open and automated harvesting and access to their collected results through REST APIs, although lack of a shared authentication strategy is a remaining challenge. With DLHub, we will soon enable execution of models and servables on distributed resources (e.g., Jetstream, Amazon Web Services, DOE Leadership Computing Facilities), and will enable the linking of servables to better represent the often-complex logic seen in machine learning applications. These and other efforts continue to move the community forward to the ultimate goal of realizing a complete data ecosystem to support the application of machine learning in materials science. In the meantime, the examples presented in this article highlight how cohesive infrastructure services, such as MDF and DLHub, can streamline complex materials discovery tasks.

## Code and Data Availability

As DLHub and MDF are both evolving projects, the code presented in this article will change over time. To best enable researchers to reproduce and build upon this work, we provide a growing repository of worked examples from various scientific domains, accessible via https://www.dlhub.org. Access to all code, packages (e.g., Forge and other clients), documentation, and interfaces related to MDF can be found via https://www.materialsdatafacility.org. Access to all code, packages (e.g., DLHub SDK and CLI), documentation, and interfaces related to DLHub can be found at https://www.dlhub.org.

## Acknowledgements

**MDF:** This work was performed under financial assistance award 70NANB14H012 from U.S. Department of Commerce, National Institute of Standards and Technology as part of the Center for Hierarchical Material Design (CHiMaD). This work was performed under the following financial assistance award 70NANB19H005 from U.S. Department of Commerce, National Institute of Standards and Technology as part of the Center for Hierarchical Materials Design (CHiMaD).This work was also supported by the National Science Foundation as part of the Midwest Big Data Hub under NSF Award Number: 1636950 "BD Spokes: SPOKE: MIDWEST: Collaborative: Integrative Materials Design (IMaD): Leverage, Innovate, and Disseminate." **DLHub:** This work was supported in part by Laboratory Directed Research and Development funding from Argonne

National Laboratory under U.S. Department of Energy under Contract DE-AC02-06CH11357. We also thank the Argonne Leadership Computing Facility for access to the PetrelKube Kubernetes cluster and Amazon Web Services for providing research credits to enable rapid service prototyping. This research used resources of the Argonne Leadership Computing Facility, a DOE Office of Science User Facility supported under Contract DE-AC02-06CH11357.

The authors would also like to acknowledge and thank the researchers who made their datasets and/or models and codes openly available [26,30,32].